\documentclass[twocolumn]{aa} 
\usepackage{savesym}
\usepackage{amsmath}
\savesymbol{iint}
\usepackage{graphicx}
\usepackage{xspace}
\usepackage{booktabs}
\usepackage[varg]{txfonts}
\usepackage{natbib,twoopt}
\usepackage[breaklinks=f]{hyperref}
\usepackage{epsfig}
\usepackage{txfonts}
\usepackage{amssymb}
\usepackage{hyperref}
\usepackage{color}
\restoresymbol{TXF}{iint}
\usepackage{tablefootnote}

\bibpunct{(}{)}{;}{a}{}{,} 
 
\newcommandtwoopt{\citeads}[3][][]{\href{http://adsabs.harvard.edu/abs/#3}%
{\citealp[#1][#2]{#3}}}
\newcommandtwoopt{\citepads}[3][][]{\href{http://adsabs.harvard.edu/abs/#3}%
{\citep[#1][#2]{#3}}}
\newcommandtwoopt{\citetads}[3][][]{\href{http://adsabs.harvard.edu/abs/#3}%
{\citet[#1][#2]{#3}}} 
\newcommandtwoopt{\citeyearads}[3][][]%
{\href{http://adsabs.harvard.edu/abs/#3}{\citeyear[#1][#2]{#3}}}

 \def\m2s2{\hbox{\,m$^{2}$\,s$^{-2}$}}       
 \def\cms2{\hbox{\,cm\,s$^{-2}$}}           
 \def\gcm3{\hbox{\,g\,cm$^{-3}$}}            
\begin{document}
   \title{No Planet around the K Giant Star 42 Draconis
\thanks{Based  in part on observations obtained at the
    2-m-Alfred Jensch Telescope at the Th\"uringer
    Landessternwarte Tautenburg}}

\author{A. P. Hatzes\inst{1}
\and
 V. Perdelwitz\inst{2}
 \and
 M. Karjalainen\inst{3}
\and
J. K\"ohler\inst{1}
\and
M. Hartmann\inst{1}
\and
M. Endl\inst{4} \\
         }

\institute{Th\"uringer Landessternwarte Tautenburg,
                Sternwarte 5, D-07778 Tautenburg, Germany \\
                 \email{artie@tls-tautenburg.de}
\and
Department of Earth \& Planetary Sciences, Weizmann Institute of Science, Rehovot 76100, Israel \\
\email{volker.perdelwitz@weizmann.ac.il}
\and
Astronomical Institute, Czech Academy of Sciences, 251 65, Ondrejov, Czech Republic
\email{marie.karjalainen@asu.cas.cz}
\and
McDonald Observatory, The University of Texas at Austin,
    Austin, TX 78712, USA
    \email{mike@astro.as.utexas.edu}
}

\date{Received; accepted}

\abstract
{Published radial velocity measurements of the K giant star 42 Dra taken over a three year  time span reveal variations consistent with a 3.9 $M_{Jup}$ mass companion in a 479-d
   orbit.}
   {This exoplanet can be confirmed if these variations are long-lived
   and coherent. Continued monitoring may also reveal additional companions.}
   {We have acquired additional radial velocity measurements of 42 Dra so that the
   data now span fifteen years. Standard periodogram analyses were used to investigate
   the stability of the planet radial velocity signal. We also investigated variations in the spectral line shapes using the bisector velocity span as well
   as infrared photometry from the COBE mission. }
   {
The recent radial velocity measurements do not follow the published planet orbit. An orbital solution using  the 2004 - 2011 data yields a period and eccentricity consistent with the published values, but the radial velocity amplitude has decreased
   by  a  factor of four from the earlier measurements. Including some additional 
radial velocity measurements taken between 2014 and 2018 reveal a second period at 530 d.
The beating of this period with the one at  479-d  may account for the observed amplitude variations.   The planet hypothesis
is conclusively ruled out by COBE/DIRBE 1.25 $\mu$ photometry that shows variations with the planet orbital period
as well as a 170 d period.}
{The radial velocity of 42 Dra shows significant amplitude variations which along with the
COBE/DIRBE photometry  firmly established that there is no giant planet around this star.
The presence of multi-periodic variations suggests that these we are seeing stellar oscillations in this
star, most likely oscillatory convection modes. These oscillations may account for some of the long period
radial velocity variations attributed to planets around K giant stars which may skew the statistics of planet
occurrence around intermediate mass stars.
 Long-term monitoring with 
excellent sampling is required to exclude amplitude variations in the long-periods
found in radial velocity of K giant stars.
}


\titlerunning{No Planet around 42  Dra}
\maketitle

\section{Introduction}

Radial velocity (RV) surveys of K-giant stars  have shown to be effective means of probing planet
formation around intermediate-mass (IM) stars with masses 1.3 to 2 $M_\odot$. 
These  measurements for IM main sequence stars are not conducive for RV planet search surveys, 
since   A to early F stars have high effective temperatures which
results in relatively  few spectral lines. These stars  also tend to have high rates of rotation. Few and broad spectral
lines  results in an  RV precision of several
tens to hundreds m\,s$^{-1}$ making the detection of sub-stellar companions difficult.
 On the other hand, intermediate-mass stars that have evolved up the giant branch
have cooler effective temperatures, thus more spectral lines, and much slower rotation
rates. They are thus highly amenable to Doppler surveys for planet searches.
There have been  several studies   been 
for planetary companions  to K giant stars with the Doppler method (e.g. Frink et al. 2002;
Setiawan et al. 2003; D\"ollinger et al.
2007; Johnson et al. 2007; Sato et al. 2008; Niedzielski et al. 2009; Wittenmyer et al.
2011; Jones et al. 2011, Lee et al. 2012). These surveys have discovered over 100 giant planets around
IM evolved stars. 

However, K giants have there own  disadvantages for Doppler surveys. First, these stars have
$p$-mode oscillations which introduce intrinsic variations in the form of RV jitter. The amplitudes
are proportional to the luminosity of the star (Kjeldsen \& Bedding 1995), so this RV jitter becomes larger as one moves
up the giant branch. For stars near the bottom of the giant branch this is as low as a few
m\,s$^{-1}$ and can be as high as tens of m\,s$^{-1}$ for a more evolved star. 

Second, they can exhibit rotational modulation due to stellar structure most likely associated
with magnetic activity. The expected rotational periods are several hundreds of
days, comparable to the orbital periods found for many stellar companions to K giant stars.
The first hint for rotational modulation induced RV variations
 was in $\alpha$ Boo. Hatzes \& Cochran (1993)  found RV variations
with a period of 231 d, nearly identical to the 233 d rotational period inferred from
He I 10830 {\AA} variations by Lambert (1987).  Finally, these stars may have new and unknown
types of oscillations. 

Unfortunately, we know very little about activity and stellar structure on slowly
rotating single  giant stars, or even the presence of long-period oscillations. To exclude the possibility of rotational modulation as
a source of the RV variability, most investigators  look for variations
 in  standard activity indicators. These include photometry,  often utilizing
the archive photometry from the astrometric space mission {\sc Hipparcos}, Ca II H \& K,
the  Ca II infrared triplet, or H$\alpha$ (e.g. Hatzes et al. 2015). 
If variations are found in any of these indicators
with the same period as the RV variations, then serious doubt is cast on the planet hypothesis.

Stellar line bisectors have also become a common tool for proving the planet hypothesis.
If any surface inhomogeneities (spots, abundance, convective velocities, etc.) are present
then these should be accompanied by variable distortions in the spectral line shapes. Indeed, it is 
these line distortions that mimic an RV variation by causing a shift in the centroid of the spectral line
as the star rotates. Line bisector measurements  were used to show that the RV
variations of HD 166435  were due spots rather than a short-period giant planet 
(Queloz et al. 2001) as well as to confirm the planet hypothesis for 51 Peg b
(Hatzes et al. 1998).

Spectral line bisector variations generally require spectra taken at very high
resolving power ($R$ $=$ $\lambda/\delta\lambda$ $\ge$ 100\,000). With data taken at lower spectral resolutions
a lack of bisector variations is a necessary condition to confirm a planet, but it is
by no means a sufficient one. A case in point is the purported planet around TW Hya.
RV measurements showed evidence for the presence of a short-period Jupiter-mass planet
around this T Tauri star (Setiawan et al. 2008). These measurements were taken at reasonably high
resolution (resolving power, $R$ = $\lambda$/$\delta\lambda$ = 45\,000). 
An observed lack of a correlation between the RV and the  spectral line bisector variations presumably 
``confirmed'' the planet. However,  subsequent RV measurements taken in the infrared showed that these
had one-third the  amplitude of the optical measurements (Hu\'elamo et al. 2008). 
The periodic RV signal was clearly due to spots.

So, the generally accepted procedure for confirming the planetary nature
of periodic RV signal is to look at as many  ancillary data as possible. If one does
not see the RV period in any of these, then the planet hypothesis is ``confirmed''. However,
for giant stars this lack of variability may lead to wrong conclusions.

In an extensive study of Aldebaran, Hatzes et al. (2015) analyzed 30 years of RV data for this
star. These seemed to show a  long-lived periodic signal at 629-d  that was interpreted
as a due to a planetary companion. There were no variations at the 629-d RV period in the equivalent widths of the
activity indicators H$\alpha$ or Ca II $\lambda$8662,  {\sc Hipparcos}  photometry or in the
spectral line shapes. An additional, intermittent signal at 520-d was seen in the RV residuals, H$\alpha$ and Ca II  and
at one-third this period in the spectral line bisectors. The planet hypothesis was the logical conclusion. However,
Reichert et al. (2019) argued against the planet hypothesis based on additional RV measurements. These
showed  that 
the $\sim$620-d period had phase shifts and there were times when it was not present at all.  

The K giant star $\gamma$ Dra also showed evidence for a ``fake'' planet. Seven years of RV
measurements showed coherent, long-lived periodic variations at 702 d consistent with an 11 $M_{Jup}$
(Hatzes et al. 2018) . The lack of variability at this  period in the Ca II S-index, spectral line shapes, 
and  {\sc Hipparcos} photometry seemed to support the planet hypothesis. In fact, the authors were
preparing a paper as a  planet discovery around $\gamma$  Dra when   more RV measurements spanning 
an additional two years became  available. These showed that the 702-d period disappeared
during 2011-2017, only to return in 2014 but with a phase shift. This behavior was reminiscent to that
seen in Aldebaran and thus refuted the planet hypothesis.

\begin{figure}[h]
\resizebox{\hsize}{!}{\includegraphics{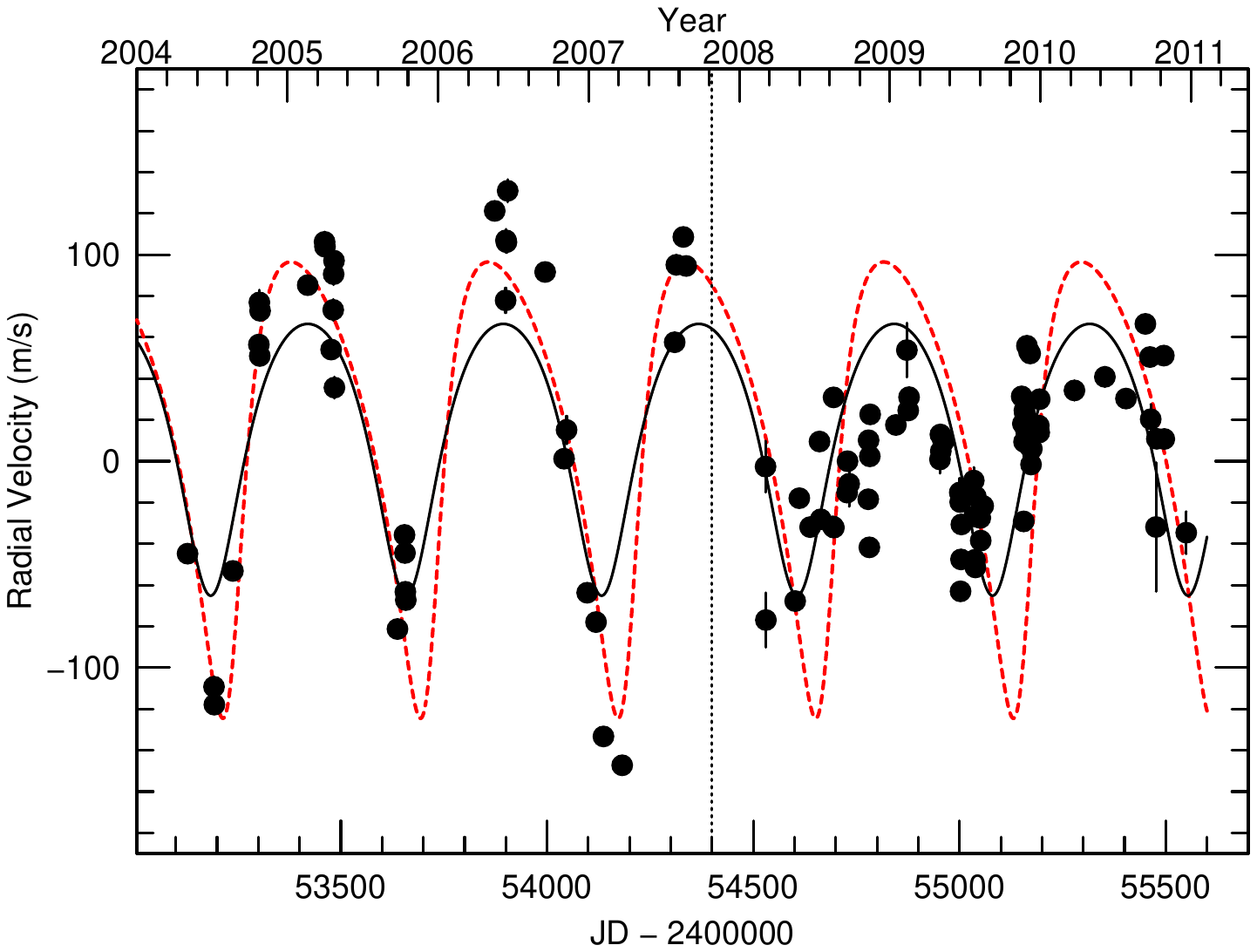}}
\caption{The RV measurements for 42 Dra. The vertical dashed line
marks the boundary between the old and new RV measurements.
The dashed curve is the orbital solution from D09 and the solid
curve a new solution based on all the 2004 $-$ 2011 data.}
\label{fig:time}
\end{figure}

At the Th\"uringer Landessternwarte Tautenburg (TLS) we have been monitoring a 
sample of K giant stars with the Doppler
method to search for exoplanet companions. This program has resulted in discoveries of the planetary
companions around 4 UMa (D\"ollinger et al. 2007), 42 Dra and  HD 139357 (D\"oellinger et al.
2009a; hereafter D09), 11 UMi, and HD 32518 (D\"ollinger et al. 2009b).  RV measurements of the giant star
42 Dra  varied with a period 479.1 d and  a $K$-amplitude of $K$ = 110.5 m\,s$^{-1}$.
This was consistent with the presence of a sub-stellar companion 
with a minimum mass of 3.9 $M_{Jup}$. {\sc Hipparcos} photometry showed no variations with the
RV period. Furthermore, the RV signal seemed to be relatively long-lived and coherent
having the same amplitude for almost three full orbital periods, or about 3.3 years. No bisector measurements
were performed on the star, but measurements of the equivalent width of
the H$\alpha$ line showed no correlation with the RV data. 
The RV variations showed all the earmarks of stemming from a planetary
companion.

\begin{figure}[h]
\resizebox{\hsize}{!}{\includegraphics{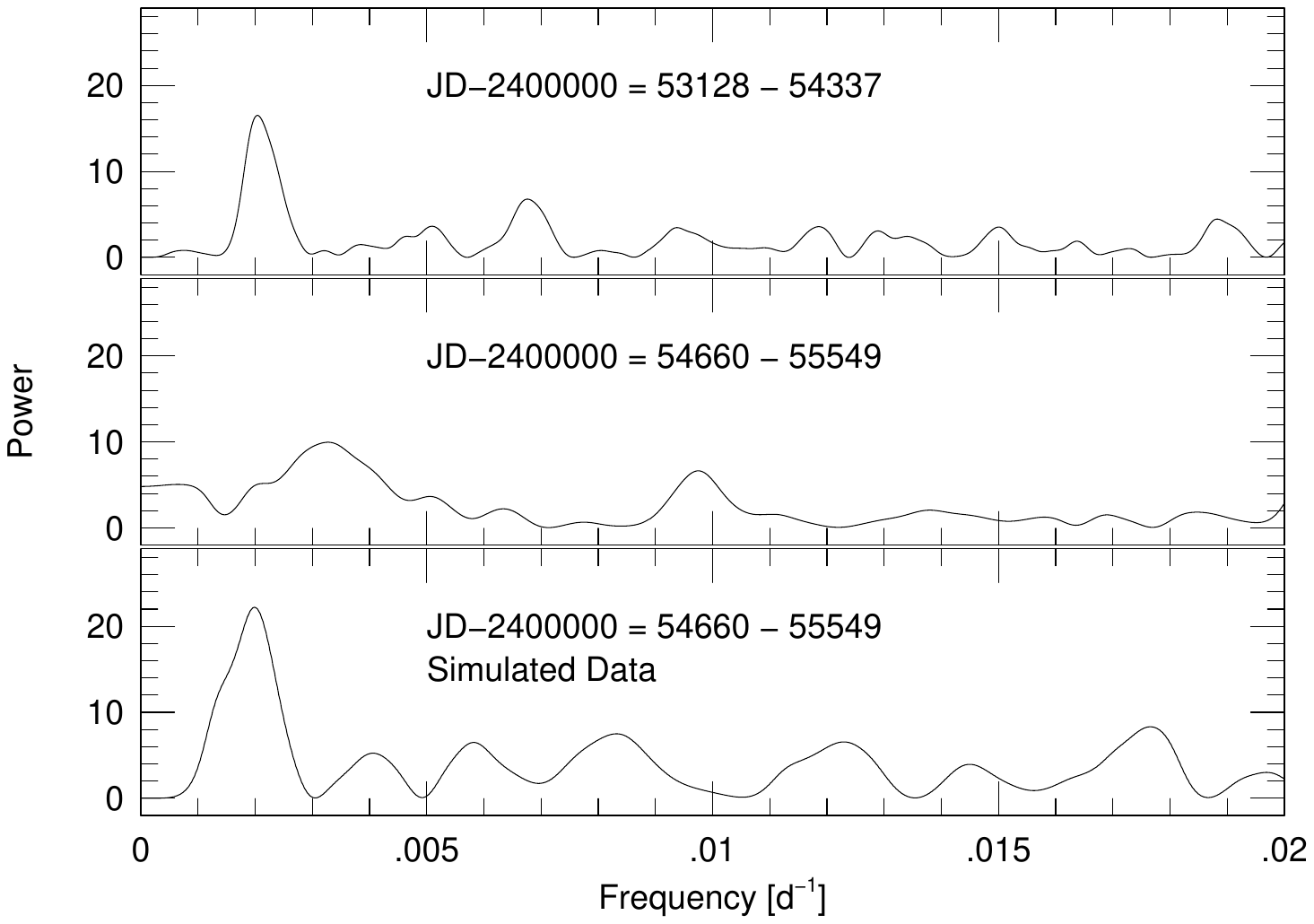}}
\caption{(Top) The Lomb-Scargle periodogram of the RV measurements up to JD = 2454337.
(Middle) The periodogram of the RV measurements taken after JD $\approx$  2454660.
(Bottom) The periodogram of a simulated orbit  over the same time range as 
the middle panel.  The 479-d
should have been present in the periodogram of the latter RV measurements. }
\label{fig:RVperiodogram}
\end{figure}

\begin{figure}[h]
\resizebox{\hsize}{!}{\includegraphics{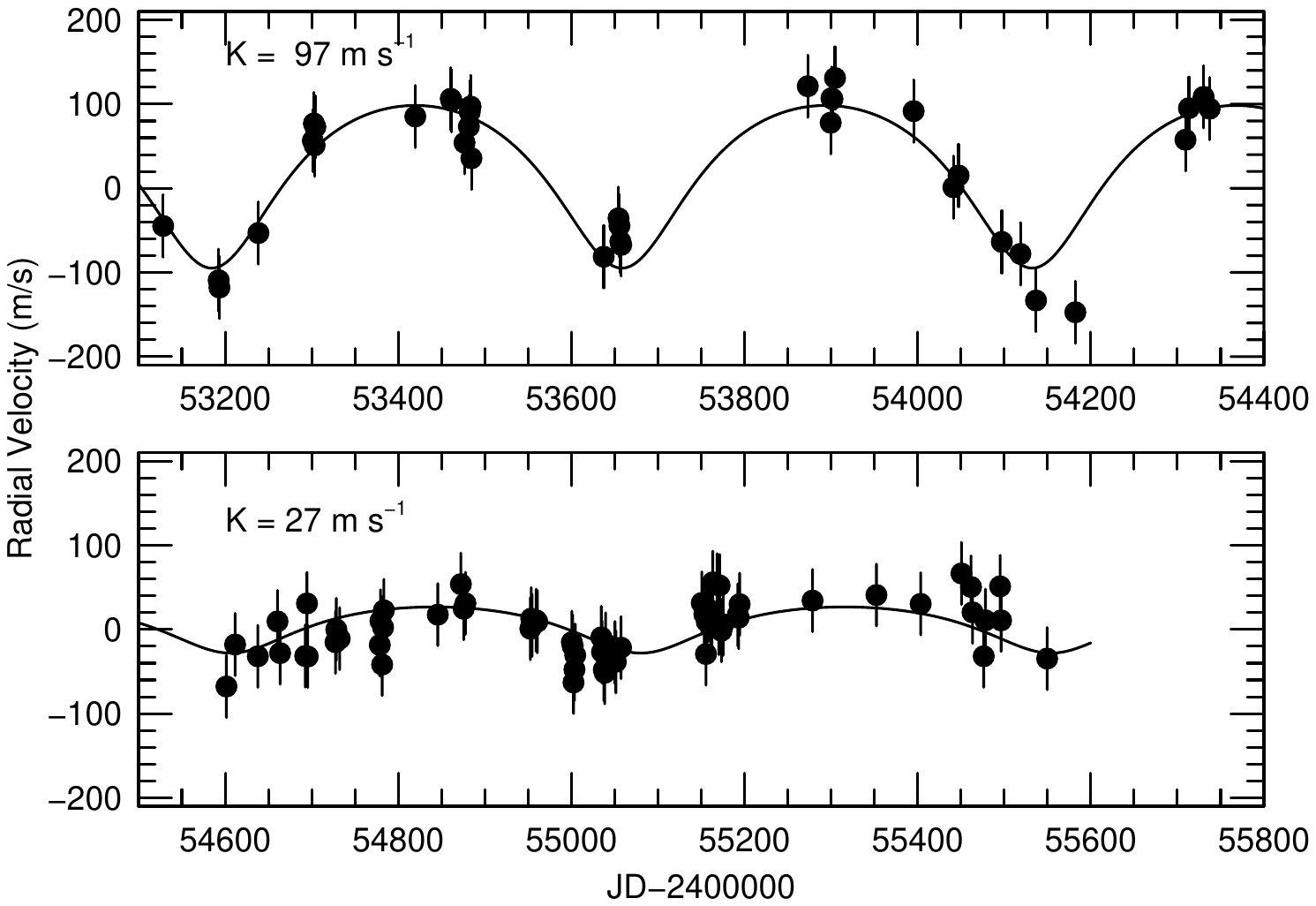}}
\caption{(Top) The orbital solution of the data up to JD $\approx$  2454400. The $K$-amplitude
is 96 m\,s$^{-1}$. 
(Bottom) An orbital solution to the RV data taken from JD $\approx$ 2454400 $-$ 2455600. All orbital
parameters except the RV amplitude were kept fixed for both data subsets.
The $K$-amplitude for the latter data is  27 m\,s$^{-1}$.}
\label{fig:kamp}
\end{figure}

We continued to monitor 42\,Dra with RV measurements for an additional nine years, three of these with good cadence. 
This was  not  only to confirm the planet hypothesis for the 479-d variations (given the experience with Aldebaran and
 $\gamma$ Dra),  but also to look for additional companions. An examination
of the  infrared photometry from the COBE mission was also made.  All these show that the RV variations seen in 42 Dra do not arise from the
 orbital reflex motion of the 
host star due to a sub-stellar companion. 

\section{Stellar parameters}
Since stellar parameters for evolved stars are often not included in catalogs like Gaia (Gaia collaboration et al. 2021) or the TESS Input Catalog (Paegert et al. 2021), we derived these values using the open-source package {\it Stellar Parameters of Giants and more} \citep{2018A&A...616A..33S}. SPOG+ uses Bayesian inference along with observational data to derive fundamental stellar properties, which are listed in Table 1  with the input parameters  and their sources.
All parameters are very well-constrained, and we found that they are in good agreement with those stated in other papers (e.g. Yu et al. 2023; Khalatyan et al. 2024).

The stellar radius from SPOG+ also agrees well with the radius measured from via interferometry. Baines et al. (2018) measured an
angular diameter of 2.048 $\pm$ 0.009 mas for 42 Dra. Using the Gaia parallax this results in a stellar radius of $R$ = 19.78 $\pm$ 0.17 $R_\odot$.
\section{Data Acquisition}

In the time span 2008 $-$ 2011 observations were made with  very good
cadence  using the Tautenburg Coude Echelle Spectrograph  (TCES)  of the 2-m
Alfred Jensch Telescope.  The spectral data covered a wavelength region of 4700 {\AA} to 7400 {\AA} at a spectral
resolving power of $R$ = 67\,000. An iodine absorption cell placed in the optical path was used to provide
the wavelength reference for the precise stellar RVs (see Hatzes et al. 2005).

Between 2014 to 2018 additional measurements were made but at much reduced cadence. These
were also taken with a different instrumental setup, namely a new echelle grating and a new CCD detector.
These data will have a different zero point offset compared to the 2008 $-$ 2011 measurements.

Table 2 lists  the RV measurements from 2004 $-$ 2011. These include    values  derived from previous published data.
Since the RV analysis was done on the
full data set values will differ from the values  published in D09.
 Table 3 is  for the RV measurements from 2014 $-$ 2018. RVs were produced with the  pipeline
{\tt  viper} (Zechmeister et al. 2021). 

Our approach is to   first examine the data taken in 2008 $-$ 
2011 when the ``planet'' signal was present to see if these give early
hints that it is not a planet. RV measurements in the final years are added to the analysis to
shed light on the true nature of the RV variations.

\section{Results}

Figure~\ref{fig:time} shows the  RV measurements from 2004 $-$ 2011  as well as the orbital solution of
D09 whose parameters
are $P$ = 479.1 $\pm$ 6.2 d, $e$ = 0.38 $\pm$ 0.06, $\omega$ =218.7 $\pm$ 10.6 degrees,
and $K$ = 110.5 $\pm$ 10.6 m\,s$^{-1}$. 
Clearly, the measurements taken from JD = 2454600 to 2455600 do not fit the orbit.

We performed an orbital solution using the 2004 - 2011  data set. The parameters
are listed in Table 4.
The period and eccentricity agree 
with the orbital solution of D09, but the full data  yield a slightly
shorter period of 474 d. (In the discussion that follows we will refer to the planet period by its original
value of 479 d). Overall,  one could still conclude that the signal
of the planetary companion was present for at least seven years. 
However, there is a discrepancy  in that the current $K$-amplitude
(65 m\,s$^{-1}$) is nearly a factor of two smaller than the earlier $K$-amplitude
of 110 m\,s$^{-1}$.

\subsection{Amplitude Variations}

 We investigated whether the high cadence data from 2004 $-$ 2011 alone were sufficient
 to provide strong evidence against the planet hypothesis.  The
  top panel of Figure~\ref{fig:RVperiodogram} shows the Lomb-Scargle (L-S) periodogram (Scargle 1982) of the data up to JD = 2454337. One sees the 
significant signal at a frequency of $\nu$ = 0.0021 d$^{-1}$, near the orbital frequency that was first reported in D09.
On the other hand, the periodogram for the RV data  from JD = 2454660 $-$ 245549 shows 
  no significant peak at the planet orbital period (central panel).  The strongest peak  actually is at $\nu$ = 0.00327 d$^{-1}$ ($P$ = 305 d) and
 is modestly significant (false alarm probability $\approx$ 0.3 \%).

We checked whether the original 479 d period could have been detected in the  RV data JD = 2454660 $-$ 245549 if it 
were indeed present.
We sampled the original orbital solution shown in Figure~\ref{fig:time} in the same manner
as the real data and added noise at a level of 55 m\,s$^{-1}$. This value
is much higher than our typical measurement error, but it is the level of the
scatter about the orbital solution. It is also consistent with the level of RV jitter expected
from stellar oscillations in this star (see below). 

The lower panel of Figure~\ref{fig:RVperiodogram} shows the periodogram of the synthetic data, but only 
using those time stamps from the middle panel. Clearly, we should have easily detected the orbital motion even when using
only the new RV measurements.

\begin{figure}[h]
\resizebox{\hsize}{!}{\includegraphics{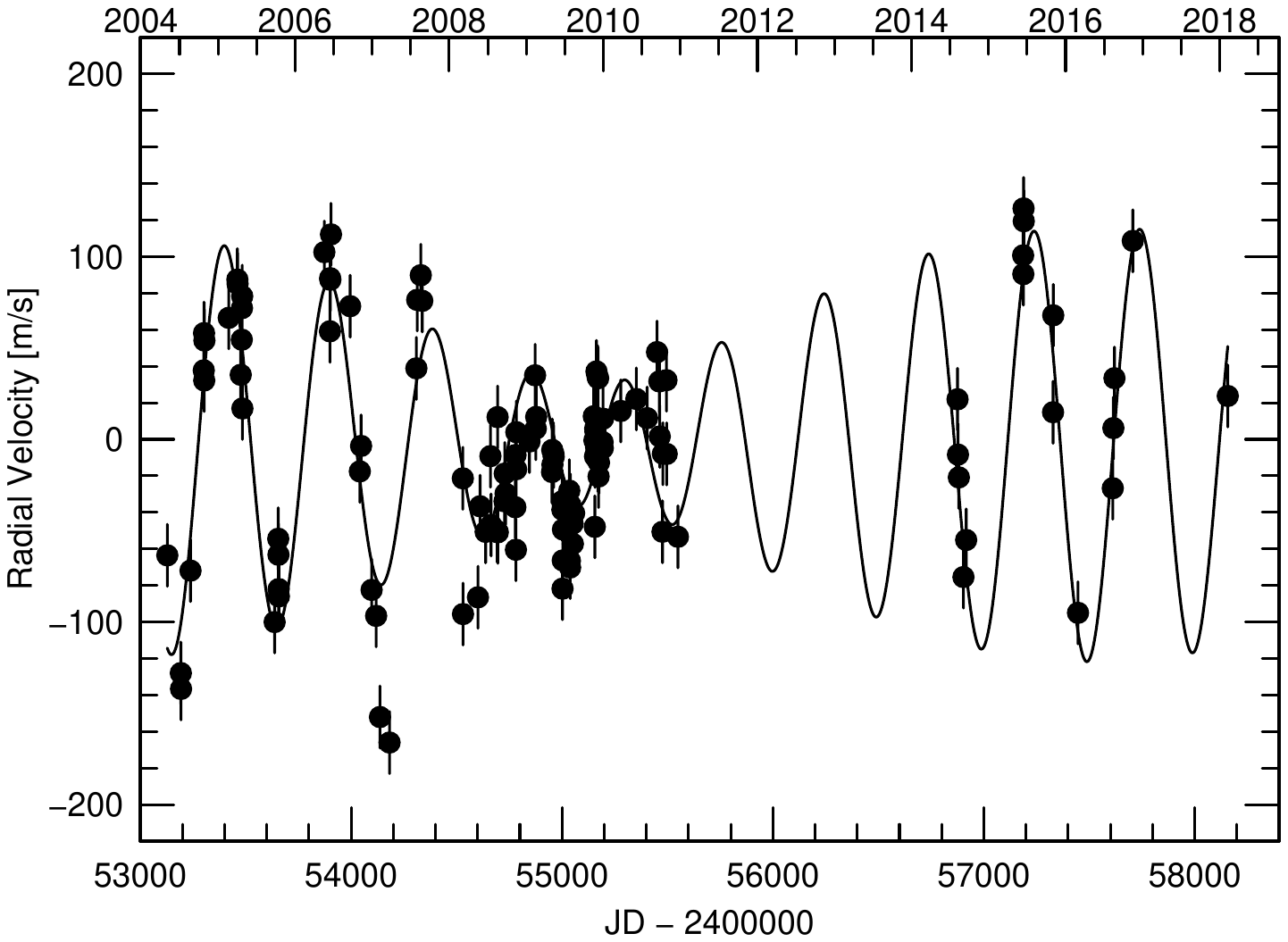}}
\caption{The complete RV measurements of 42 Dra from 2004 $-$ 2018 (points). The curve
represents a two-component fit with two periods, $P_1$ = 487.3 d and $P_2$ = 530 d. }
\label{fig:2perfit}
\end{figure}

The RV data show a clear variation in the amplitude of the 479-d signal.  We took our
revised orbital solution and used the  parameters to fit the RV data up to JD = 2454530.
All parameters were kept fixed and only the $K$-amplitude was allowed to vary. 
This resulted in  $K$ = 96.6 $\pm$ 6.3 m\,s$^{-1}$ (upper panel of Fig.~\ref{fig:kamp}).
We then fit the RV data taken after JD = 245660
again allowing only
the RV amplitude to vary. (We removed two data points in the fit in order
to provide a larger time gap between the two subset data.) This resulted in $K$ = 27.4 $\pm$ 6.1 m\,s$^{-1}$
(lower panel of Fig.~\ref{fig:kamp}). If the 479-d period were present, its  RV amplitude has been reduced by about a factor
of four from the earlier measurements. This is inconsistent with 
the orbital reflex motion of the host star due to a companion.

\subsection{Frequency Analysis of the Full Data Set}

We then performed a frequency analysis of the full RV data set covering JD = 2453128  $-$ 2458156 (2004 $-$ 2018)
that is  shown in Figure~\ref{fig:2perfit}.
This was done using a pre-whitening procedure. Periodic signals were sequentially found and subtracted and
 a search for additional signals was made on the residuals. The process was stopped when the final peak had a false alarm probability, FAP $<$ 0.01.
 The FAP was estimated from the height of the peak compared to the surrounding noise level (i.e. the signal-to-noise ratio, SNR). This SNR  can be
 converted into a FAP (Kuschnig et al. 1997; Hatzes 2019).
Three significant frequencies were found, shown in Table 5.
The curve in Figure~\ref{fig:2perfit} shows a fit to the RV data using the first two periods (487-d and 530-d) which better shows the beating between these two dominant periods.

\subsection{An Examination of the Indicators for Planet Confirmation}

As discussed in the introduction, it is wise to look for variations with the RV period
in other quantities. If these show variations with the same
period as the RV, then the planet hypothesis can be rejected.
D09 examined the equivalent width of H$\alpha$ in 42 Dra and found no 
variations with the RV period, but the authors did not investigate any line
shape variations.  Here we take a closer look at the spectral line shapes as well as {\sc Hipparcos} photometry.

\subsubsection{Spectral Line Shape Variations}

To investigate changes in the spectral line shapes we  calculated the cross-correlation
function (CCF) using  the spectral region 4790--4900 {\AA} and the first
observation as a template. This region 
is largely free of the iodine absorption lines that we use for our wavelength calibration.  
We then calculated the bisector of the CCF -- the locus of the midpoints calculated from both sides of the
CCF having the same flux value. A linear least-squares fit was then made to the
CCF bisectors. We finally converted this slope  between the CCF height values of 0.3 to 0.85 to an equivalent velocity which we will call the bisector velocity span (BVS).

The top panel of Fig.~\ref{fig:hipp_bvs} shows the L-S periodogram of the BVS measurements for 2004 to 2011 when
the planet signal was present. There is a modest peak
at a frequency of $\nu$ = 0.001497 $\pm$ 0.000060 d$^{-1}$ ($P$ = 671 $\pm$ 28 d) with an amplitude
of $\approx$ 60 m\,s$^{-1}$. This frequency is significantly different from the orbital
frequency of the purported planet. A bootstrap analysis (see below) indicates
a relatively low FAP of $\approx$ 5 $\times$ 10$^{-4}$. The crucial point is that the BVS frequency 
does not  coincide with the orbital frequency. 

\subsubsection{{\sc\bf Hipparcos} photometry revisited}

\begin{figure}[h]
\resizebox{\hsize}{!}{\includegraphics{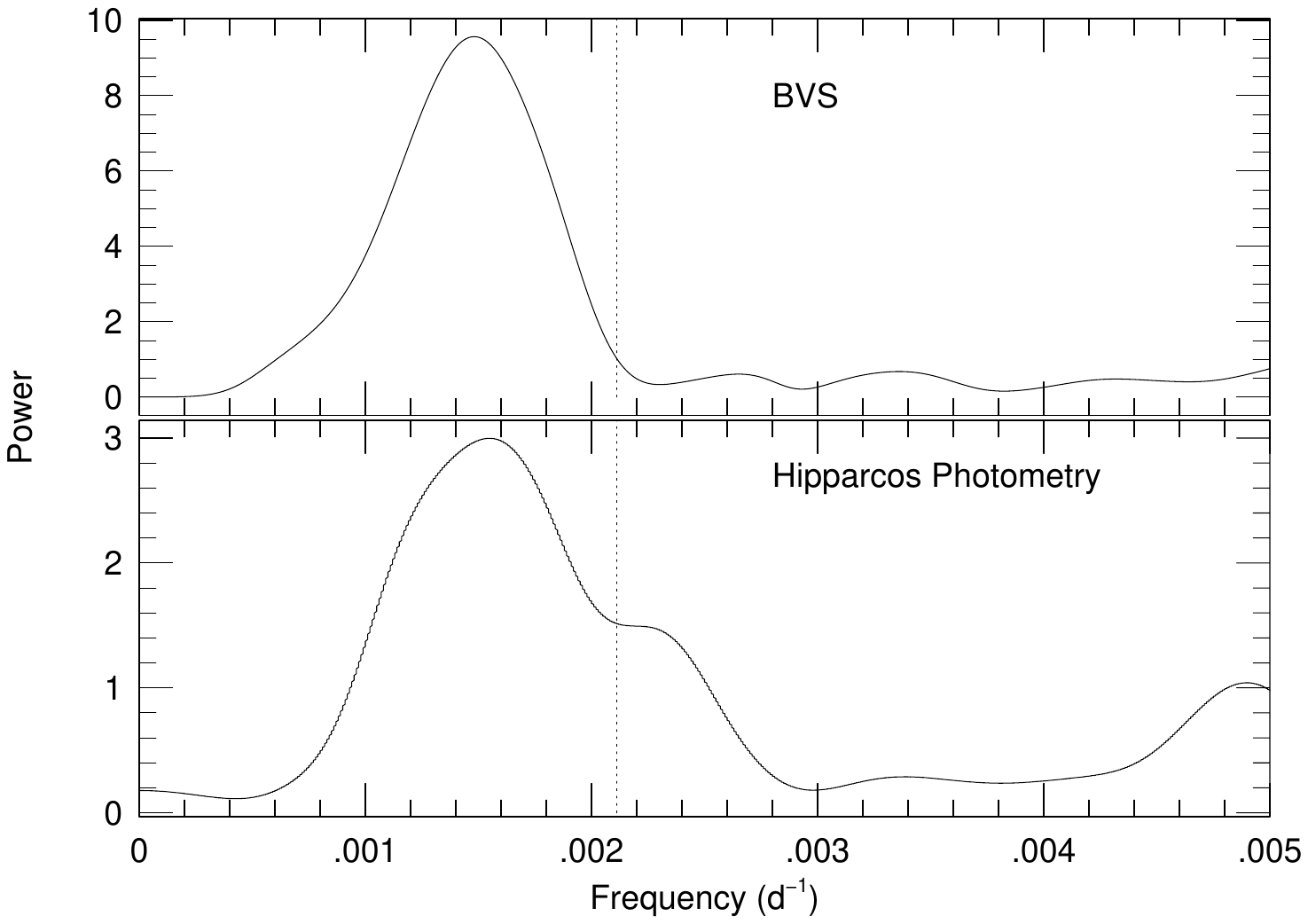}}
\caption{(Top) The Lomb-Scargle periodogram of the bisector velocity span (BVS) during the
time when the planet signal was present. (Bottom) The Lomb-Scargle periodogram
of the {\sc Hipparcos} photometry. We should note that the photometry was not contemporaneous with the BVS measurements. The
dashed vertical line is the orbital frequency of the purported planet.}
\label{fig:hipp_bvs}
\end{figure}

D09 showed that there were no significant
variations in the {\sc Hipparcos} 
photometry with the same period as the RV variations. 
We took another look at the {\sc Hipparcos} photometry to see if there were any low
frequency signals in the data that could possibly coincide with the BVS variations.
The lower panel of Fig.~\ref{fig:hipp_bvs} shows the low frequency end of L-S periodogram
of the   {\sc Hipparcos} photometry. There is indeed  a peak at $\nu$ = 0.00145 $\pm$ 0.00018 
d$^{-1}$ ($P$ = 690 $\pm$ 90 d) that is consistent  with the one seen
in the BVS, but the signal is not very  significant (FAP $\sim$ 0.05). The {\sc Hipparcos} photometry does
not refute the planet hypothesis.

\subsection{DIRBE Photometry: Final Refutation of the Planet Hypothesis}

\begin{figure}[h]
\resizebox{\hsize}{!}{\includegraphics{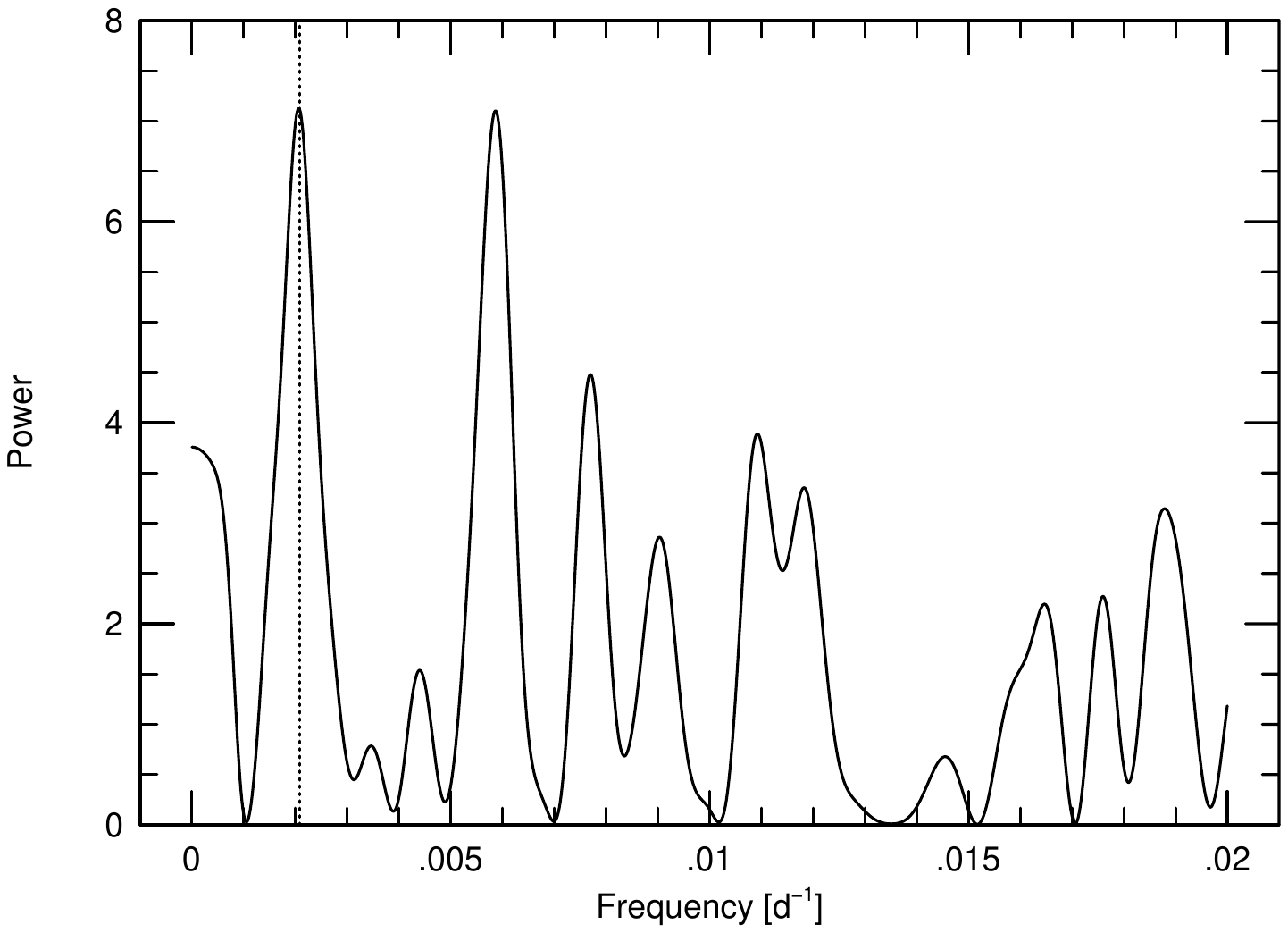}}
\caption{Lomb-Scargle periodogram of the DIRBE 1.25 $\mu$m photometry. The
dashed vertical line is the orbital frequency of the purported planet.}
\label{fig:dirbe_LS}
\end{figure}

\begin{figure}[h]
\resizebox{\hsize}{!}{\includegraphics{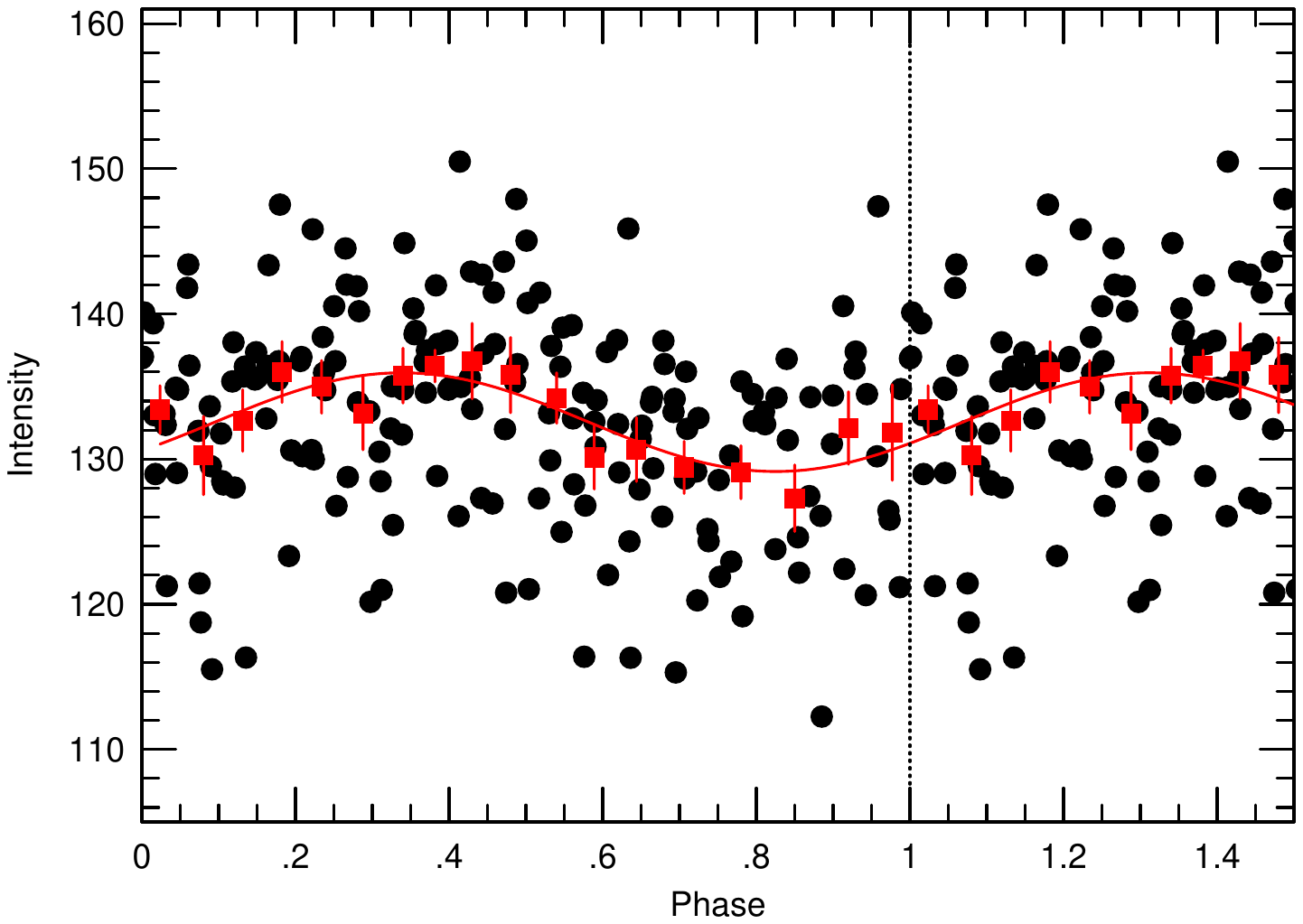}}
\caption{The DIRBE photometry phased to the 479-d ``orbital'' period. The
red squares are phase-binned values. The curve represents a sine fit to the
data. Data measurements are repeated to the right of the vertical dashed line.
}
\label{fig:dirbe_phase}
\end{figure}

Price et al. (2010) extracted weekly average near infrared fluxes for 2652 stars in the
all-sky maps of the Diffuse Infrared Background Experiment on the Cosmic
Background Explorer (COBE/DIRBE); hereafter ``DIRBE photometry''). We used the time series of 1.25 $\mu$m fluxes for
42 Dra to search for periodic signals.
Figure~\ref{fig:dirbe_LS} shows the L-S periodogram of the DIRBE photometry. The strongest peak
coincides with the orbital frequency of the planet shown by the vertical dashed line. There is 
    a second strong peak at $\nu$ = 0.00589 d$^{-1}$ ($P$ = 169.8 d). Figure~\ref{fig:dirbe_phase} shows the DIRBE photometry phased to the 479-d period.

We assessed the false alarm probability (FAP) using the bootstrap method (Murdoch et al. 1993).
In this method the RV data are randomly
shuffled a large number (10$^{5}$)  of times keeping the time stamps fixed.
The fraction of random power larger than the data power yields the FAP. Since we are interested in 
the FAP at a known frequency in the data we employed a 
 ``windowing'' bootstrap (Hatzes 2019). The FAP is determined via bootstrap
over a modest sized window centered on the frequency of interest. The window is then successively narrowed and a new
FAP calculated. The final FAP is the extrapolated value at  zero width for the window  at the
frequency of interest. This resulted in
a  FAP $\approx$ 2  $\times$ 10$^{-3}$ that random noise could produce the observed power {\it exactly} at the 
planet orbital frequency. 

We also assessed the FAP of the second peak using the bootstrap, but over a larger frequency range $\nu$ = 0 $-$ 0.02 d$^{-1}$ since
the 169.8-d  period is not a known signal. After  removing the contribution
of the 479-d signal we found a    FAP  $\approx$ 0.01 that a peak would have higher power anywhere in the
frequency range of interest.

\subsection{Short-term RV variations}

\begin{figure}[h]
\resizebox{\hsize}{!}{\includegraphics{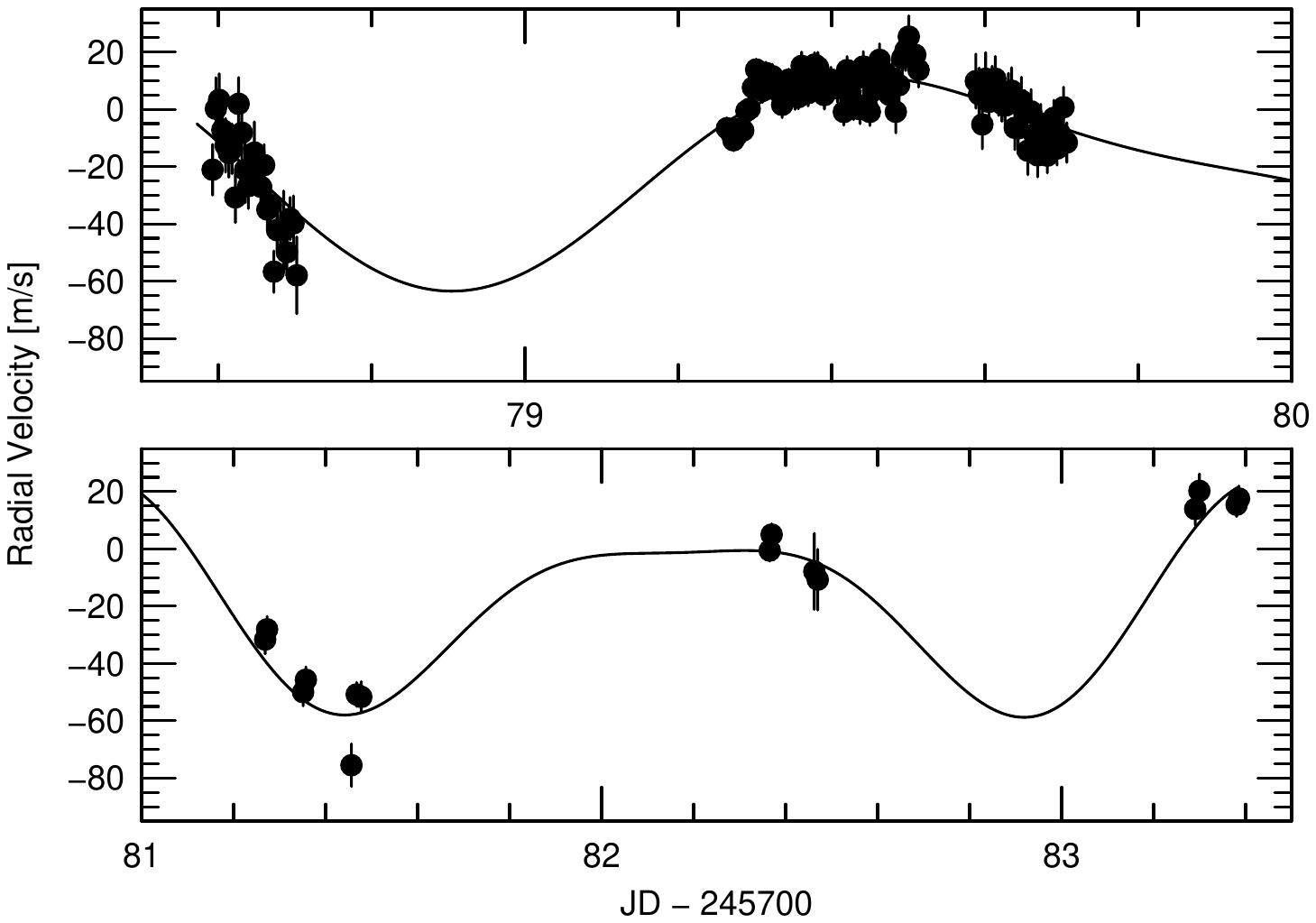}}
\caption{Short term RV variability of 42 Dra over several nights. The curve is a two-period
fit with period, $P_1$ = 1.3 d, amplitude $K_1$ = 79.3 m\,s$^{-1}$; $P$ = 0.82 d, 
$K_2$ = 49.2 m\,s$^{-1}$. }
\label{fig:shortRV}
\end{figure}

As part of our study of 42 Dra we investigated the short term RV variability by observing the
star with high cadence over several nights. Figure~\ref{fig:shortRV} shows the
best time series from these observations (RV values are listed in Table 6).

It is beyond the scope of this paper  for a detailed study of the short term variations - our
observations are too sparse. Nevertheless, we performed a frequency analysis
to assess the rough time scales and amplitudes involved. This resulted in two dominant
signals, one with frequency $\nu_1$ = 0.768 d$^{-1}$ ($P_1$ = 1.3 d) and RV amplitude $K_1$ = 79.3 m\,s$^{-1}$
and $\nu_2$ = 1.21  d$^{-1}$ ($P$ = 0.82 d) and RV amplitude $K_2$ = 49.2 m\,s$^{-1}$. The curve
in the figure shows the two-component fit.

Clearly, these are p-mode oscillations. Kjeldsen \& Bedding (1995) gave simple scaling relations
for the frequency of  maximum power, $\nu_{max}$, and amplitudes for these stellar oscillations.
Using the stellar parameters of 42 Dra, these scaling relationships give 
$\nu_{max}$  $\approx$ 0.88 $d^{-1}$ ($P$ = 1.14 d), close to our value for $\nu_1$. 
These  relationships also give a predicted velocity amplitude of $\approx$ 30 m\,s$^{-1}$.

Kjeldsen \& Bedding (2011) gave revised scaling relations for the oscillation amplitude taking into
account the stellar temperature and with a stronger dependence on the stellar mass. These 
require a knowledge of the mode lifetimes which are not well known for an evolved K giant star
like 42 Dra. Assuming a mode lifetime comparable to the sun ($\approx$ 2.9 days) results in a 
velocity amplitude of $\approx$ 15 m\,s$^{-1}$. However, red giant stars can have mode lifetimes of up to a month
(De Ridder et al. 2009) or even more for a giant like 42 Dra. This results in a predicted oscillation amplitude of  $\approx$ 50 m\,s$^{-1}$.
Thus the observed amplitudes for the short term variations in 42 Dra are entirely consistent with
stellar oscillations.

\section{Discussion}

Precise RV measurements of 42 Dra taken between 2004 and 2008 reveal variations with a period
of 479 d and an amplitude of 110 m\,s$^{-1}$ that were attributed to a planetary companion (D09). 
This detection passed all the standard tests for planet confirmation. There were no H$\alpha$ variations indicative of activity. 
{\sc Hipparcos} photometry and line bisectors showed variations with a period of 690-d that were well
separated from the orbital period of the purported planet. Furthermore, the 479-d variations were coherent
and present for at least four years. As a planet detection, 42 Dra b was as solid as many other claimed planets.
The original conclusion by D09 seemed reasonable.

Additional RV monitoring of this star tells  another story. RV measurements taken between
2008 and 2011 show that the $K$-amplitude decreased abruptly from about
100 m\,s$^{-1}$ to 27 m\,s$^{-1}$. Furthermore, if one only considers the RV data from
2008 to 2011, a peak in the periodogram appears at the wrong period. Simulations using
the published orbital solution sampled in the same manner as the real data indicate that we should have found the planet
signal in the subset data.  Even in the 2008 $-$ 2011 data  there were indications that the planet was
not real which was finally confirmed by the  DIRBE photometry that showed significant periodic variations exactly at  the orbital 
period of the planet

A frequency analysis of the full data set spanning 2004 $-$ 2018 reveal a possible origin
of the amplitude variations. This results in an additional period at  530 d. The beating of these
two closely spaced periods can mimic amplitude variations (Fig.~\ref{fig:2perfit}). It is impossible
for these two periods to be planetary companions as this would require giant planets in nearly the same orbit. This system would be dynamically unstable.

It is beyond the scope of this paper to perform a detailed dynamical analysis. However, the stability can be estimated by 
the Hill radius, $r_H$, given by:
$$r_H \approx a \left( {m} \over {3M}\right)^{1/3}$$
where $a$ and $m$ are  the semi-major axis and mass of the smaller body, and $M$ the mass of the star.
A minimum separation of approximately 3.5 $R_H$ is widely accepted as the minimum requirement 
for long-term stability. Using the amplitudes and periods from Table 5, our two hypothetical
planets would have masses of approximately 3.2 $M_{Jup}$ and 2  $M_{Jup}$ and semi-major
axes of 1.24 AU and 1.32 AU, respectively. Each planet has a Hill radius $r_H$ $\approx$ 0.1, comparable to the minimum
separation. This suggests an unstable system.

Gladman (1993) also established the stability of a three body system based on the  Hill criterion.
Consider a  two-planet system  with mass $m_1$ and $m_2$ in orbit around a star of mass $M$. 
We can denote the semi-major axis
of the outer planet by $a$ = 1 $+$ $\Delta$, where $\Delta$ is the fractional separation,
If the  mass ratios    of the two planets  with respect to the host star are  given by  $\mu_1$ and  $\mu_2$ then
orbits are most likely stable  if $$\Delta   >  2.4( \mu_1 + \mu_2)^{1/3}$$
For the hypothetical planets of 42 Dra, $\Delta$ = 0.08, $\mu_1$ = 2.8 $\times$ 10$^{-3}$ and $\mu_2$ = 1.9  $\times$ 10$^{-3}$.
In this case $\Delta$ is less than 2.4($\mu_1 + \mu_2$)$^{1/3}$ $=$ 0.4, so the system is unstable.

It is not clear what the nature of the 479-d RV period is, the one initially attributed
to a companion. One candidate would be rotational modulation by surface features. 
So far at least  five periods have been identified in 42 Dra: $\approx$ 480 d (RV and DIRBE photometry),  535 d  (RV), 294 d (RV),
690 d ({\sc Hipparcos} photometry and BVS) and 170 d (DIRBE photometry). These are not harmonics of each other which is typical for
rotational modulation,  so only one can be the rotation period. The fact that the 479-d period is also
seen in the DIRBE photometry suggests that this may be due to rotational modulation by surface features. However, photometric variations were also seen at 690-d in the {\sc Hipparcos} photometry, albeit with lower significance.

One can estimate the rotational period from the  radius and the rotational velocity
of the star. Jofre et al. (2015) measured a projected rotational velocity $v$\,sin$i$ = 1.76 $\pm$ 0.45 km\,s$^{-1}$ which
yields a maximum rotation period of 554 $\pm$  142 d. This is consistent with either the 479-d period or the 
690-d period found in the {\sc Hipparcos} photometry and the BVS. We cannot be certain which of these is the
rotation period.

Given the large number of periods which may be present in the RV and photometry of 42~Dra,
the simplest explanation is that these
arise from stellar oscillations. Long period RV variations with comparable periods have been found
in other evolved stars. 
 For example, Jorissen et al. (2016) found RV variations with  a K-amplitude of
540 m\,s$^{-1}$ and a period P = 285 d in the Carbon-enhanced metal poor (CEMP) star
HE 0017+0055. Since variability  with similar periods were found in other CEMP stars, they
proposed that the RV variations may be due to envelope oscillations.
 Oscillatory convective modes  were suggested for  the long-period RV variations in $\gamma$ Dra (Hatzes et al. 2018).
 Only additional studies, both observational and theoretical are needed to confirm the type of oscillations
 seen in 42 Dra.  

42 Dra is yet another case where all the standard tools of planet confirmation of an RV discovery
seem to have failed.
This only emphasizes that when confirming such long period RV variations in K giant
stars one needs  to look not only at as many ancillary measurements as 
possible (which may still fail), but  also to take measurements spanning more than a decade 
to search for amplitude variations and closely spaced periods. 
Long-term monitoring of K giants
with planet candidates may be essential for confirmation of the planet hypothesis.
Additionally, one can use RV measurements taken at infrared wavelengths to confirm
planet detections around K giant stars (Trifonov et al. 2015).

Long-period RV variations attributable to stellar oscillations have been found in several
K giant stars. This raises the question: ``{\it How many planets around evolved stars are false?''}.
The ``fake'' planets around evolved
IM stars may skew the statistics on the frequency of giant planet formation 
around more stars more massive than the sun which could have implications for planet formation theories.

It is worth  noting that producing ``real'' planets also has  implications in public outreach.
Recently, the International Astronomical Union (IAU) has started to assign proper names
to exoplanets that have been discovered, many  of these with the RV method. The planet around
42~Dra was assigned the name {\it Orbitar}\footnote{nameexoworlds.iau.org} which now 
would have to be retracted. This is not the first planet to be refuted, nor the last.
Besides K giant stars this has happened with the
``disappearance'' of planets around the main sequence stars GL 581 (Forveille et al. 2011;
Robertson et al. 2014) and $\alpha$ Cen B (Hatzes 2013; Rajpaul et al. 2016). The
 3.2 $M_\oplus$ planet claimed around Barnard's Star (Ribas et al. 2018) was later
 shown to be a false positive due to the one-year alias of an stellar activity signal (Lubin et al. 2021).

We welcome the effort to broaden the appeal of the exciting field of exoplanet research
to the public and 42~Dra b ({\it Orbitar})  shows the public that  science is not always
perfect. Scientists sometimes get it wrong in pursuit of the truth.  There will continue to 
be cases  where forms of intrinsic variability (sometimes unknown)  can mimic a planet signal. 
Astronomers and the public should realize that many RV planet discoveries are still candidates
that must be confirmed by independent methods. We note that the Gaia mission can be 
provide astrometric observations that could confirm the planetary nature of the long period RV
variations in K giant stars.
 A naming  convention is commendable, but  it is more important that the IAU 
 adopt a set of criteria that  exoplanet discoveries have to pass before they are declared as bona 
fide planets.  The complex RV variability of K giants only highlights our ignorance
regarding stellar phenomena in these stars. Continued studies of these stars are required
to sort out which variations are companions, and which are due to intrinsic stellar
variability and these may require dedicated surveys lasting ten years of more.

\begin{acknowledgements}
This research has made use of the SIMBAD database operated
at CDS, Strasbourg, France. APH acknowledges the support of DFG grants HA 3279/5-1 and HA 3279/8-1.
\end{acknowledgements}

\clearpage

\begin{table}
\centering
\begin{tabular}{l|c}
Parameter & Value \\\hline\hline
mag$_B$$^1$ & $6.005\pm0.014$~mag \\
mag$_V$$^2$\   & $4.823\pm0.009$~mag \\
parallax$^3$    & $11.056\pm0.0841$~ mas \\
$[$Fe/H$]^4$    &  $-0.4326\pm0.01$~dex \\
 E(B-V)$^5$  & $0.05\pm0.01$~mag \\
 A$_V$$^6$   & $0.17\pm0.01$~mag \\ \hline
 M & $1.07\pm0.01$~M$_{\odot}$\\
 R & $19.17^{+0.3}_{-0.17}$~R$_{\odot}$\\ 
 log~$g$ & $1.9\pm0.02$~cm/s$^2$\\
log($\tau$ [yr]) & $9.81\pm0.04$\\
L & $129.26^{+1.77}_{-1.05}$~L$_{\odot}$\\ 
T & $4449.08^{+11.40}_{-16.72}$~K \\\hline
\end{tabular}
\caption{\label{tab:params}Stellar parameters for 42 Dra.  The input values for SPOG+ (above the line) are taken from: (1, 2) H{\o}g et al. (2000),  (3) Gaia collaboration et al. (2021), (4) Abdurro'uf et al (2022), (5, 6) Gontcharov \& Mosenkov (2017). Those parameters derived with SPOG+ are listed below the line.
}
\end{table}

\begin{table}
\begin{center}
\begin{tabular}{lrr}
Julian Day   & RV (m\,s$^{-1}$)  & $\sigma$ (m\,s$^{-1}$) \\
\hline
2453128.4179  &  -63.53  &  4.39 \\
2453192.5234  &  -128.00  &  4.87 \\ 
2453193.4023  &  -136.67  &  3.99 \\
2453238.4101  &  -71.90  &  3.22 \\
2453301.3281  &  37.73  &  4.10 \\
2453302.5234  &  58.13  &  6.00 \\
2453303.4179  &  32.20  &  4.25 \\
2453304.2578  &  54.08  &  3.74 \\
2453419.7109  &  66.50  &  3.70 \\
2453460.6171  &  87.46  &  4.78 \\
2453461.6367  &  85.11  &  4.02 \\
2453476.5234  &  35.30  &  3.49 \\
2453481.5664  &  54.53  &  5.09 \\
2453482.5742  &  71.82  &  5.13 \\
2453483.5078  &  78.34  &  4.31 \\
2453484.5273  &  16.88  &  5.19 \\
2453637.3281  &  -100.05  &  3.61 \\
2453654.4453  &  -54.45  &  3.83 \\
2453655.5195  &  -63.21  &  2.82 \\
2453656.5390  &  -82.01  &  2.78 \\
2453657.5234  &  -86.01  &  3.05 \\
2453899.5312  &  59.13  &  6.01 \\
2453900.4804  &  88.22  &  5.40 \\
2453901.5703  &  87.32  &  5.05 \\
2453904.5039  &  112.20  &  5.43 \\
2453873.0468  &  102.46  &  3.37 \\
2453995.3398  &  72.88  &  3.78 \\
2454041.2890  &  -17.56  &  3.41 \\
2454047.2656  &  -3.62  &  6.70 \\
2454097.2226  &  -82.45  &  4.04 \\
2454136.6992  &  -152.06  &  4.05 \\
2454118.6835  &  -96.67  &  2.49 \\
2454182.1562  &  -166.03  &  2.92 \\
2454309.3828  &  38.85  &  4.06 \\
2454313.3750  &  76.34  &  4.69 \\
2454330.3281  &  89.84  &  4.36 \\
2454337.3359  &  75.80  &  3.73 \\
2454660.4882  &  -9.22  &  1.86 \\
2454663.5000  &  -46.89  &  1.80 \\
2454692.5156  &  -50.31  &  2.19 \\
2454694.5312  &  12.14  &  2.22 \\
2454695.4218  &  -50.92  &  2.37 \\
2454637.7500  &  -50.68  &  3.84 \\
2454529.6445  &  -21.38  &  12.39 \\
2454530.5234  &  -95.68  &  13.20 \\
2454601.4921  &  -86.51  &  2.60 \\
2454611.5273  &  -36.67  &  3.58 \\
2454727.5312  &  -33.96  &  3.78 \\
2454728.4335  &  -18.75  &  3.42 \\
2454732.3359  &  -29.71  &  10.96 \\
2454778.6562  &  -37.18  &  3.00 \\
2454779.4492  &  -8.61  &  3.60 \\
2454781.3789  &  -60.49  &  2.93 \\ 
\hline \\
\end{tabular}
\caption{RV measurements for 42 Dra 2004 - 2011 }
\end{center}
\label{tab:RV1}
\end{table}

\begin{table}
\setcounter{table}{1}
\begin{center}
\begin{tabular}{lrr}
Julian Day   & RV (m\,s$^{-1}$)  & $\sigma$ (m\,s$^{-1}$) \\
\hline
2454782.4140  &  -16.40  &  3.92 \\
2454783.3398  &  3.92  &  2.51 \\
2454845.6640  &  -1.20  &  4.05 \\
2454872.3359  &  35.04  &  13.11 \\
2454875.5664  &  5.80  &  2.19 \\
2454877.5898  &  12.25  &  1.53 \\
2454952.5898  &  -17.91  &  6.70 \\
2454953.5859  &  -5.82  &  3.41 \\
2454954.5820  &  -13.88  &  3.78 \\
2454959.3281  &  -7.92  &  5.74 \\
2454960.5703  &  -9.52  &  3.97 \\
2455000.4726  &  -33.91  &  6.84 \\
2455001.4882  &  -38.54 &  3.40 \\
2455002.4257  &  -81.78  &  3.20 \\
2455003.5351  &  -66.32  &  2.97 \\
2455004.4140  &  -49.38  &  3.33 \\
2455034.5390  &  -28.14  &  6.40 \\
2455035.3789  &  -45.51  &  3.99 \\
2455037.4257  &  -66.52  &  4.17 \\
2455038.4843  &  -70.20  &  2.61 \\
2455039.5468  &  -35.91  &  2.82 \\
2455050.3437  &  -46.22  &  3.33 \\
2455051.3554  &  -57.22  &  4.37 \\
2455057.3437  &  -40.42  &  2.52 \\
2455150.6015  &  12.63  &  2.81 \\
2455153.5625  &  -0.59  &  3.63 \\
2455155.4687  &  -47.92  &  2.74 \\
2455156.3437  &  -9.39  &  9.03 \\
2455157.2968  &  5.85  &  3.57 \\
2455158.4179  &  4.11  &  1.89 \\
2455161.2656  &  -4.12  &  4.27 \\
2455162.2851  &  -7.01  &  3.53 \\
2455163.2890  &  37.15  &  4.27 \\
2455168.2968  &  34.32  &  2.66 \\
2455170.3203  &  -11.73  &  3.98 \\
2455171.2226  &  33.40  &  3.63 \\
2455173.2890  &  -20.36  &  2.86 \\
2455175.3046  &  -12.73  &  2.43 \\
2455192.1835  &  -1.52  &  3.10 \\
2455193.2460  &  -4.85  &  4.85 \\
2455194.3242  &  11.27  &  2.63 \\
2455278.5546  &  15.51  &  4.72 \\
2455352.3710  &  22.08  &  4.74 \\
2455403.4765  &  11.57  &  3.84 \\
2455450.5898  &  47.74  &  3.84 \\
2455461.6171  &  31.69  &  3.93 \\
2455463.4257  &  1.56  &  7.160 \\
2455478.2578  &  -7.91  &  4.03 \\
2455495.2421  &  32.37  &  3.77 \\
2455496.3789  &  -8.00  &  3.84 \\
2455549.5898  &  -53.38  &  10.20 \\ 
\hline \\
\end{tabular}
\caption{RV measurements for 42 Dra 2004 - 2011 (cont.) }
\end{center}
\end{table}

\begin{table}
\begin{center}
\begin{tabular}{lrr}
Julian Day   & RV (m\,s$^{-1}$)  & $\sigma$ (m\,s$^{-1}$) \\
\hline
2456875.4094 & 21.91  & 2.28 \\
2456877.5337 & -8.38  & 5.04 \\
2456881.5527 & -20.85  & 3.54 \\
2456903.4301 & -75.42  & 2.37 \\
2456916.2838 & -55.13  & 4.52 \\
2457186.4182 & 100.69  & 4.72 \\
2457187.4240 & 90.43  & 4.60 \\
2457188.4137 & 126.42  & 4.08 \\
2457189.4683 & 119.40  & 5.00  \\
2457328.3674 & 14.82  & 6.07  \\
2457329.2581 & 67.88  & 7.10  \\
2457446.7115 & -95.02  & 4.21  \\
2457611.5435 & -26.78  & 4.31 \\
2457614.3892 & 6.14  & 2.92 \\
2457619.4522 & 33.46 & 5.32 \\
2457706.4234 & 108.65 & 5.21 \\
2458156.7328 & 23.72 & 4.25 \\
\hline \\
\end{tabular}
\caption{RV measurements for 42 Dra 2014 - 2018 }
\end{center}
\label{tab:RV2}
\end{table}

\begin{table}
\begin{center}
\begin{tabular}{lc}
Parameter  & Value     \\
\hline
Period [days]                    & 473.9  $\pm$ 4.0            \\
T$_{0}$ [JD-2450000]             & 2711.60 $\pm$ 32.7           \\
$K$ [m\,s$^{-1}$]                & 65.8 $\pm$  5.9            \\
$e$                              & 0.25 $\pm$ 0.007            \\
$\omega$ [deg]                   & 181.2 $\pm$ 19.8           \\
$f(m)$ [solar masses]            & (1.27 $\pm$ 0.34) $\times$ 10$^{-8}$   \\
$m$ sin $i$ [$M_{Jupiter}$]      & 2.45 $\pm$ 0.21                           \\
$a$ [AU]                         & 1.18 $\pm$0.06                              \\
\hline
\end{tabular}
\caption{Orbital solution of the purported planet around 42 Dra using the full RV data.
}
\end{center}
\label{tab:orbitparm}
\end{table}

\begin{table}
\begin{center}
\begin{tabular}{ccccc}   
  Period                      & Amplitude                 &   Phase                     & SNR  & FAP \\
  (d)                          &   (m\,s$^{-1}$)          &                                   &           &      \\
\hline
   487.8  $\pm$ 0.9   &  79.3  $\pm$ 3.5  &  0.75 $\pm$ 0.01  & 6.9        &   5.3 $\times$ 10$^{-18}$ \\
    534.8   $\pm$  2.2   & 49.2  $\pm$ 3.5  & 0.39  $\pm$ 0.01  & 5.3         &  1.0 $\times$  10$^{-8}$ \\
     294.1     $\pm$ 0.2   & 20.7  $\pm$ 3.5  & 0.40  $\pm$ 0.03  &  4.7          &  3.7 $\times$ 10$^{-6}$ \\                                                                
\hline
\end{tabular}
\caption{Frequencies found in the  full RV data set.
}
\end{center}
\label{tab:frequencies}
\end{table}

\begin{table}
\begin{center}
\begin{tabular}{lrr}
Julian Day   & RV (m\,s$^{-1}$)  & $\sigma$ (m\,s$^{-1}$) \\
\hline
2454778.5926 & -21.08 & 8.86 \\
2454778.5968 & 0.17 & 10.87 \\
2454778.6011 & 3.30 & 9.11 \\
2454778.6053 & -7.21 & 9.36 \\
2454778.6095 & -12.59 & 9.41 \\
2454778.6138 & -15.28 & 8.48 \\
2454778.6180 & -13.76 & 8.72 \\
2454778.6222 & -30.75 & 8.80 \\
2454778.6265 & 1.91 & 9.21 \\
2454778.6307 & -8.17 & 9.17 \\
\hline 
\end{tabular}
\caption{Short-term RV measurements for 42 Dra (full table in electronic form)}
\end{center}
\label{tab:RVshort}
\end{table}

\end{document}